\documentclass[prb,showpacs]{revtex4}

\usepackage{graphicx}
\usepackage{dcolumn}
\usepackage{bm}

\begin{document}

\title{Static Solitons of the Sine-Gordon Equation and Equilibrium Vortex
Structure in Josephson Junctions}
\author{S. V. Kuplevakhsky}
\email{kuplevakhsky@ilt.kharkov.ua}
\author{A. M. Glukhov}
\affiliation{B. I. Verkin Institute for Low Temperature Physics and Engineering, \\
National Academy of Sciences of Ukraine, \\
47 Lenin Ave., 61103 Kharkov, UKRAINE}
\date{\today}

\begin{abstract}
The problem of vortex structure in a single Josephson junction in an
external magnetic field, in the absence of transport currents, is
reconsidered from a new mathematical point of view. In particular, we derive
a complete set of exact analytical solutions representing all the stationary
points (minima and saddle-points) of the relevant Gibbs free-energy
functional. The type of these solutions is determined by explicit evaluation
of the second variation of the Gibbs free-energy functional. The stable
(physical) solutions minimizing the Gibbs free-energy functional form an
infinite set and are labelled by a topological number $N_v=0,1,2,\ldots $.
Mathematically, they can be interpreted as nontrivial ''vacuum'' ($N_v=0$)
and static topological solitons ($N_v=1,2,\ldots $) of the sine-Gordon
equation for the phase difference in a finite spatial interval: solutions of
this kind were not considered in previous literature. Physically, they
represent the Meissner state ($N_v=0$) and Josephson vortices ($%
N_v=1,2,\ldots $). Major properties of the new physical solutions are
thoroughly discussed. An exact, closed-form analytical expression for the
Gibbs free energy is derived and analyzed numerically. Unstable
(saddle-point) solutions are also classified and discussed.
\end{abstract}

\maketitle

\section{Introduction}

In this paper, we reconsider the old physical problem\cite{J65,KY72,BP82} of
equilibrium vortex structure in a single Josephson junction from a new
mathematical point of view. The necessity of such reconsideration is
motivated by the fact that, in spite of significant contribution by a number
of authors (see, e.g., Refs. \cite{K66,OS67,Ga84,Yu94,Se04}), the problem
did not find an exact and complete analytical solution in previous
theoretical literature. In particular, the basic question, namely,
\textquotedblright What should be called an equilibrium Josephson vortex in
precise mathematical terms?\textquotedblright , remained unanswered. Over
years, this issue has been a source of considerable misunderstanding. For
example, there still exists a wide-spread erroneous belief\cite{BCG92,Kr01}
that Josephson vortices \textquotedblright do not form\textquotedblright\ in
\textquotedblright small\textquotedblright\ junctions with $W\ll 2\lambda
_{J}$, where $W\,$ is the length of the insulating barrier and $\lambda _{J}$
is the Josephson length.\cite{J65,KY72,BP82} (In reality, Josephson vortices
do form for arbitrary small $W>0$, provided the externally applied magnetic
field $H$ is sufficiently high: see section V of this paper.)

To clarify the situation, we consider the simplest case of a junction in a
constant, homogeneous external magnetic field, in the absence of externally
applied currents. Relevant geometry in presented in Fig. 1. In particular,
the $x$ axis is perpendicular to the insulating layer (the barrier); the $y$
axis is along the barrier. The barrier length $W=2L$ is assumed to be
arbitrary: $0<W<\infty $. A constant, homogeneous external magnetic field $%
\mathbf{H}$ is applied along the axis $z$: $\mathbf{H}=\left( 0,0,H\geq
0\right) $. Full homogeneity along the $z$ axis is assumed.

\begin{figure}[tbp]
\includegraphics{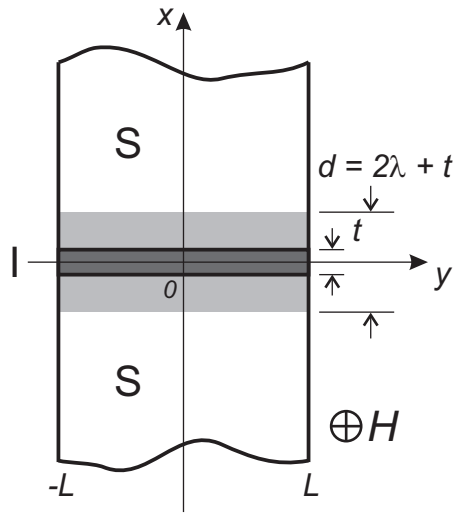}
\caption{The geometry of the problem: $t$ is the thickness of the barrier; $%
W=2L$ is the length of the barrier; $\protect\lambda $ is the London
penetration depth; $d=2\protect\lambda +t$ is the width of the
field-penetration region (shaded). The external magnetic field $H$ is
directed into the plane of the figure.}
\label{fig:junction}
\end{figure}

The difficulties of most previous approaches to the problem arose from the
incompleteness of a traditional formulation\cite{KY72,BP82,K66,OS67} in
terms of a static (time-independent) sine-Gordon equation
\begin{equation}
\frac{d^{2}\phi }{dy^{2}}=\frac{1}{\lambda _{J}^{2}}\sin \phi   \label{0.1}
\end{equation}%
for the phase difference $\phi =\phi \left( y\right) $ and physical boundary
conditions
\begin{equation}
\frac{d\phi }{dy}\left( \pm L\right) =2edH\quad \left( \hbar =c=1\right) .
\label{0.2}
\end{equation}%
Indeed, equations (\ref{0.1}), (\ref{0.2}) constitute an ill-posed
boundary-value problem\cite{CH} in the sense that they do not meet the
requirement of the uniqueness of the solution. [This is a consequence of
nonlinearity of Eq. (\ref{0.1}) and of the physical boundary conditions
being imposed on $\frac{d\phi }{dy}$ instead of $\phi $.] Thus, although the
general solution to (\ref{0.1}) was well-known,\cite{A70} the constants of
integration specifying all physically observable configurations could not be
unambiguously determined from boundary conditions (\ref{0.2}) alone.

An obvious way to overcome the problem of incompleteness lies in an
observation that Eqs. (\ref{0.1}), (\ref{0.2}) are nothing but stationarity
conditions of a relevant Gibbs free-energy functional.\cite{K04,K05}
Therefore, among all solutions to (\ref{0.1}) that satisfy (\ref{0.2}) at a
given $H$, one has to choose those that minimize (at least locally) the
Gibbs free-energy functional. By general physical arguments, the stable (in
a mathematical sense) solutions should represent an observable
thermodynamically stable state (in the case of an absolute minimum), as well
as observable thermodynamically metastable states (in the case of local
minima). Unfortunately, the issue of exact minimization of the Gibbs
free-energy functional did not receive appropriate attention in the previous
literature. (However, stability of some particular solutions was analyzed
from a somewhat different point of view, e.g., in Refs. \cite{Ga84,Yu94,Se04}%
.)

The fact that a complete set of stable solutions to (\ref{0.1}), (\ref{0.2})
can be obtained in a closed analytical form has been first demonstrated in
Refs. \cite{K04,K05} within the framework of minimization of a certain wide
class of energy functionals whose particular representative is the Gibbs
free-energy functional of a single Josephson junction. Exact analytical
solutions of Refs. \cite{K04,K05} form an infinite set and are labelled by a
topological (vortex) number $N_v=0,1,2,\ldots $. As is pointed out in Refs.
\cite{K04,K05}, these solutions constitute a new class of particular
solutions to the sine-Gordon equation that can be interpreted as nontrivial
''vacuum'' (for $N_v=0$) and static topological solitons (for $%
N_v=1,2,\ldots $) in a finite spatial interval. Solutions of this type were
not considered in the previous literature on the sine-Gordon equation.\cite%
{L80,DEGM82,KK89}

In what follows, we show that the new topological solutions resolve the
problem formulated at the beginning of the Introduction. To make the paper
self-contained and independent of Refs. \cite{K04,K05}, we employ a new,
more general method of the derivation of topological solutions. In
particular, we start with a full set of solutions to (\ref{0.1}), (\ref{0.2}%
). In contrast to the approach of Refs. \cite{K04,K05}, the selection of
stable (physical) solutions is made by explicit evaluation of the sign of
the second variation of the Gibbs free-energy functional, which allows us to
establish general conditions of stability.

In section II, we introduce the Gibbs free-energy functional and show that
all its stationary points are either minima or saddle points. In section
III, the second variation of the Gibbs free energy is discussed. We show
that the sign of the second variation is related to the sign of the lowest
eigenvalue of a certain Sturm-Liouville eigenvalue problem. In this way, we
establish general conditions of stability. In section IV, a full set of
solutions to (\ref{0.1}), (\ref{0.2}) is obtained and classified with
respect to stability. In section V, a full set of stable (physical)
solutions to (\ref{0.1}), (\ref{0.2}) is derived. These solutions comprise a
Meissner solution ($N_v=0$) and vortex solutions ($N_v=1,2,\ldots $). Their
major physical and mathematical properties are thoroughly discussed. An
exact, closed-form expression for the Gibbs free energy is derived. This
expression is analyzed numerically in the cases of a ''large'' junction and
a ''small'' junction. In section VI, unstable (saddle-point) solutions to (%
\ref{0.1}), (\ref{0.2}) are classified and discussed. In section VII, we
summarize the main results of the paper and make a few concluding remarks.
Finally, in appendix A, we analyze stability in two singular cases (the
Meissner solution in a semiinfinite interval and the single-soliton solution
in an infinite interval).

\section{Gibbs free-energy functional: Major properties}

From now on, we will employ dimensionless units. Thus, the length scale
along the $y$ axis is normalized to the Josephson penetration length $%
\lambda _J$. The magnetic field is normalized to the superheating field of
the Meissner state in a semi-infinite junction\cite{KY72} $H_s=\left(
ed\lambda _J\right) ^{-1}$. The energy scale is normalized to $\frac{%
d\lambda _JH_s^2}{16}$. (In particular, the flux quantum in our
dimensionless units is given by $\Phi _0=\pi $.)

In terms of the dimensionless units, a phase-dependent part of the Gibbs
free-energy functional per unit length along the $z$ axis takes the form\cite%
{K04,K05}
\begin{equation}  \label{1.1}
\Omega _G\left[ \phi ,\frac{d\phi }{dy};H\right] =2H^2W+\int\limits_{-L}^Ldy%
\left[ 1-\cos \phi \left( y\right) +\frac 12\left[ \frac{d\phi \left(
y\right) }{dy}\right] ^2\right] -2H\left[ \phi \left( L\right) -\phi \left(
-L\right) \right] .
\end{equation}
Note that the last term in Eq. (\ref{1.1}) is, physically, proportional to
the magnetic (Josephson) flux
\begin{equation}  \label{1.2}
\Phi _J=\frac 12\left[ \phi \left( L\right) -\phi \left( -L\right) \right] ,
\end{equation}
with
\begin{equation}  \label{1.3}
h\left( y\right) =\frac 12\frac{d\phi }{dy}
\end{equation}
being the corresponding local magnetic field (equilibrium or not).

The stationarity condition of the Gibbs free-energy functional (\ref{1.1}),
\begin{equation}  \label{1.4}
\delta \Omega _G\left[ \phi ,\frac{d\phi }{dy};H\right] =0,
\end{equation}
reduces to the static sine-Gordon equation for the phase difference
\begin{equation}  \label{1.5}
\frac{d^2\phi }{dy^2}=\sin \phi
\end{equation}
and boundary conditions
\begin{equation}  \label{1.6}
\frac{d\phi }{dy}\left( \pm L\right) =2H\geq 0.
\end{equation}
[Compare with Eqs. (\ref{0.1}) and (\ref{0.2}), respectively, where
dimensional units are employed.]

The main property of the functional (\ref{1.1}) follows from the fact that
it belongs to the class of regular functionals, i.e., satisfies a necessary
condition of the minimum;\cite{A62} hence, all stationary points of (\ref%
{1.1}) are either minima of saddle points. [Unfortunately, the statement of
Ref. \cite{Yu94} that all stationary points of (\ref{1.1}) ''are minima'' is
incorrect.]

As already mentioned in the Introduction, boundary conditions (\ref{1.6}) do
not ensure the uniqueness of the solution to Eq. (\ref{1.5}). Moreover,
solutions to (\ref{1.5}), (\ref{1.6}), as stationary points of the Gibbs
free-energy functional (\ref{1.1}), can correspond to either minima or
saddle points of this functional. Taking into account that only stable phase
configurations, corresponding to minima of (\ref{1.1}), are physically
observable, the problem one has to solve can be formulated as follows: Find
all solutions to (\ref{1.5}), (\ref{1.6}) that minimize (at least locally) (%
\ref{1.1}) at given $H\geq 0$. To resolve this problem, we have to turn to
sufficient conditions of the minimum,\cite{A62} which requires an analysis
of the second variation of (\ref{1.1}).

\section{Second variation of the Gibbs free-energy functional and a related
Sturm-Liouville eigenvalue problem}

Let $\phi =\bar \phi \left( y\right) $ be a particular solution to (\ref{1.5}%
), (\ref{1.6}) [i.e., a certain stationary point of (\ref{1.1})]. The
increment of the Gibbs free-energy functional (\ref{1.1}) in the vicinity of
this solution, induced by variations $\bar \phi \left( y\right) \rightarrow
\bar \phi \left( y\right) +\delta \phi \left( y\right) $, where $\delta \phi
$ has a continuous derivative that obeys boundary conditions
\begin{equation}  \label{2.1}
\frac{d\delta \phi }{dy}\left( -L\right) =\frac{d\delta \phi }{dy}\left(
L\right) =0,
\end{equation}
can be expanded in an infinite series:\cite{A62}%
\begin{equation}  \label{2.2}
\Delta \Omega _G\left[ \delta \phi ,\frac{d\delta \phi }{dy};H\right] _{\phi
=\bar \phi } =\frac 1{2!}\delta ^2\Omega _G\left[ \delta \phi ,\frac{d\delta
\phi }{dy}\right] _{\phi _n=\bar \phi _n}+\sum_{k\geq 3}\frac 1{k!}\delta
^k\Omega _G\left[ \delta \phi \right] _{\phi =\bar \phi }.
\end{equation}
Here,
\begin{equation}  \label{2.3}
\delta ^2\Omega _G\left[ \delta \phi ,\frac{d\delta \phi }{dy}\right] _{\phi
=\bar \phi } =\int\limits_{-L}^Ldy\left[ \cos \bar \phi \left( y\right) %
\left[ \delta \phi \left( y\right) \right] ^2+\left[ \frac{d\delta \phi }{dy}%
\left( y\right) \right] ^2\right] ,
\end{equation}
and
\begin{equation}  \label{2.4}
\delta ^k\Omega _G\left[ \delta \phi \right] _{\phi =\bar \phi }
=-\int\limits_{-L}^Ldy\cos \left[ \bar \phi \left( y\right) + \frac{k\pi }2%
\right] \left[ \delta \phi \left( y\right) \right] ^k,\quad k\geq 3.
\end{equation}

The sign of the second variation (\ref{2.3}) in the expansion (\ref{2.2})
determines a type of the tested solution $\phi =\bar \phi \left( y\right) $.
Three different cases are possible: (i) the case
\begin{equation}  \label{2.5}
\delta ^2\Omega _G\left[ \delta \phi ,\frac{d\delta \phi }{dy}\right] _{\phi
=\bar \phi }>0
\end{equation}
corresponds to a minimum of (\ref{1.1}), (i.e., the solution is inside a
stability region); (ii) the case
\begin{equation}  \label{2.6}
\delta ^2\Omega _G\left[ \delta \phi ,\frac{d\delta \phi }{dy}\right] _{\phi
=\bar \phi }\geq 0,
\end{equation}
corresponds to a boundary of the stability region (a bifurcation point),\cite%
{Th82} when the solution can loose stability with respect to a certain
variation; (iii) finally, if $\delta ^2\Omega _G\left[ \delta \phi ,\frac{%
d\delta \phi }{dy}\right] _{\phi =\bar \phi }$ has no definite sign, the
solution corresponds to a saddle point of (\ref{1.1}), which means absolute
instability. Therefore, in what follows, we concentrate ourselves on the
evaluation of $\delta ^2\Omega _G$.

As is shown in variational theory of eigenvalues,\cite{LL50} the functional $%
\delta ^2\Omega _G$ satisfies the following general relation:\cite{r1}
\begin{equation}  \label{2.7}
\delta ^2\Omega _G\left[ \delta \phi ,\frac{d\delta \phi }{dy}\right] _{\phi
=\bar \phi }\geq \mu _0\int\limits_{-L}^Ldy\left[ \delta \phi \left(
y\right) \right] ^2,
\end{equation}
where $\mu _0$ is the lowest eigenvalue of the Sturm-Liouville eigenvalue
problem
\begin{equation}  \label{2.8}
-\frac{d^2\psi }{dy^2}+\cos \vec \phi \left( y\right) \psi =\mu \psi ,\quad
y\in \left[ -L,L\right] ,
\end{equation}
\begin{equation}  \label{2.9}
\frac{d\psi }{dy}\left( -L\right) =\frac{d\psi }{dy}\left( L\right) =0.
\end{equation}
Equality on the right-hand side of relation (\ref{2.7}) is achieved, when $%
\delta \phi $ coincides (up to a factor) with the eigenfunction
corresponding to $\mu _0$: i.e., $\delta \phi \equiv const\,\psi _0$. As is
clear from relation (\ref{2.7}), the three different types (i)-(iii) of the
behavior of $\delta ^2\Omega _G$ correspond, respectively to: (i) $\mu _0>0$%
; (ii) $\mu _0=0$; (iii) $\mu _0<0$.

The properties of the self-adjoint operator, specified by the left-hand side
of Eq. (\ref{2.8}) and boundary conditions (\ref{2.9}), are well-known.\cite%
{LS70} In particular, its spectrum $\mu _n$ ($n=0,1,2,\ldots $) is discrete,
real, infinite and bounded from below:
\begin{equation}  \label{2.10}
\mu _0<\mu _1<\mu _2<\ldots
\end{equation}
(Note that in our case $\left| \mu _0\right| \leq 1$.) The corresponding
eigenfunctions $\psi _n$ ($n=0,1,2,\ldots $) are real, mutually orthogonal
and can be normalized:
\begin{equation}  \label{2.11}
\int\limits_{-L}^Ldy\psi _n\left( y\right) \psi _m\left( y\right) =0,\quad
n\neq m;\quad 0<\int\limits_{-L}^Ldy\left[ \psi _n\left( y\right) \right]
^2<\infty .
\end{equation}
The eigenvalue number $n$ determines the number of nodes of the
eigenfunction $\psi _n$ in the interval $\left[ -L,L\right] $ . Thus, the
eigenfunction $\psi _0$ can be considered strictly positive:
\begin{equation}  \label{2.12}
\psi _0=\psi _0\left( y\right) >0,\quad y\in \left[ -L,L\right] .
\end{equation}
Significantly, the set of eigenfunctions $\left\{ \psi _n\right\}
_{n=0}^\infty $ is complete in the sense that any function $f\left( y\right)
$ that possesses continuous derivatives up to the second order and obeys
boundary conditions $f\left( \pm L\right) =0$ can be expanded in terms of $%
\psi _n.$

Upon substitution of a solution of the boundary-value problem (\ref{1.5}), (%
\ref{1.6}) into (\ref{2.8}), this equation can be transformed into the
well-known\cite{WW27} Lam\'e equation. Therefore, the boundary-value problem
for $\mu =\mu _0$ can, in principle, be solved analytically. However, since
we need to know only the sign of $\mu _0$, we will derive below a set of
exact relations that will allow us to determine the sign of $\mu _0$ without
explicit evaluation of $\mu _0$.

First, we notice that the function
\begin{equation}  \label{2.13}
\chi \equiv \frac{d\phi }{dy},
\end{equation}
where $\phi =\phi \left( y\right) $ is a particular solution to (\ref{1.5}),
(\ref{1.6}) (in what follows, we drop the bar over $\vec \phi $), satisfies
the equation
\begin{equation}  \label{2.14}
-\frac{d^2\chi }{dy^2}+\cos \phi \left( y\right) \chi =0.
\end{equation}
Combining (\ref{2.13}), (\ref{2.14}) and (\ref{2.8}) (with $\mu =\mu _0$, $%
\psi =\psi _0$), using boundary conditions (\ref{2.9}), we obtain the sought
relations%
\begin{equation}  \label{2.15}
\mu _0=\frac{\psi _0\left( L\right) \frac{d^2\phi }{dy^2}\left( L\right)
-\psi _0\left( -L\right) \frac{d^2\phi }{dy^2}\left( -L\right) }{%
\int\limits_{-L}^Ldy\psi _0\left( y\right) \frac{d\phi }{dy}\left( y\right) }
=\frac{\psi _0\left( L\right) \sin \phi \left( L\right) -\psi _0\left(
-L\right) \sin \phi \left( -L\right) }{\int\limits_{-L}^Ldy\psi _0\left(
y\right) \frac{d\phi }{dy}\left( y\right) }.
\end{equation}
[For definiteness, we choose the sign of $\psi _0$ according to (\ref{2.12}%
).] In addition, we note that, if $\psi _0$ is known explicitly, $\mu _0$
can be found from the relation
\begin{equation}  \label{2.16}
\mu _0=\frac{\int\limits_{-L}^Ldy\psi _0\left( y\right) \cos \phi \left(
y\right) }{\int\limits_{-L}^Ldy\psi _0\left( y\right) }.
\end{equation}

\section{Complete set of solutions and their classification with respect to
stability}

We begin with the first integral of Eq. (\ref{1.5}):
\begin{equation}  \label{3.1}
\frac 12\left[ \frac{d\phi }{dy}\right] ^2+\cos \phi =C,\quad -1\leq
C<\infty ,
\end{equation}
where $C\,$ is the constant of integration. Using (\ref{3.1}), it is
straightforward to derive the general solution to (\ref{1.5}).\cite{A70} We
write down this solution in the following explicit form:

(I) $-1\leq C\leq 1$:
\begin{equation}  \label{3.2}
\phi _{\pm }\left( y\right) =\pi \left( 2n+1\right) \pm 2\arcsin \left[
k\,sn\left( y-y_0,k\right) \right] ,\quad n=0,\pm 1,\ldots ,
\end{equation}
\[
-K\left( k\right) \leq y_0<K\left( k\right) ,\quad k\equiv \frac{1+C}2,\quad
0\leq k\leq 1;
\]

(II) $1\leq C<\infty $:
\begin{equation}  \label{3.3}
\phi _{\pm }\left( y\right) =\pi \left( 2n+1\right) \pm 2\,am\left( \frac{%
y-y_0}k,k\right) ,\quad n=0,\pm 1,\ldots ,
\end{equation}
\[
-kK\left( k\right) \leq y_0<kK\left( k\right) ,\quad k\equiv \frac
2{1+C},\quad 0<k\leq 1.
\]
Here, the functions $am\,u$ and $sn\,u$ are the Jacobi elliptic amplitude
and the elliptic sine, respectively; $K\left( k\right) $ is the complete
elliptic integral of the first kind.\cite{AS65} The choice of sign (i.e., $%
\phi _{+}$ or $\phi _{-}$) and allowed values of the constants of
integration $y_0$, $k$ in (\ref{3.2}), (\ref{3.3}) should be determined from
the requirement of compatibility with boundary conditions (\ref{1.6}) that
we rewrite in a somewhat generalized form:
\begin{equation}  \label{3.4}
\frac{d\phi }{dy}\left( -L\right) =const\geq 0,\quad \frac{d\phi }{dy}\left(
L\right) =const\geq 0;
\end{equation}
\begin{equation}  \label{3.5}
\frac{d\phi }{dy}\left( -L\right) =\frac{d\phi }{dy}\left( L\right) .
\end{equation}

\subsection{Solutions of type I}

Consider solutions of type I [Eqs. (\ref{3.2})]. First, we note that, for
this type of solutions,
\begin{equation}
\left\vert \phi \left( L\right) -\phi \left( -L\right) \right\vert <2\pi
\label{3.5.1}
\end{equation}%
for any $0<L<\infty $. Taking into account that
\begin{equation}
\frac{d\phi _{\pm }}{dy}=\pm 2k\,cn\,\left( y-y_{0},k\right) ,  \label{3.6}
\end{equation}%
where $\,cn\,u$ is the elliptic cosine, from $\,$(\ref{3.4}) we find that
the appropriate solution is $\phi =\phi _{+}$. If $L\leq \frac{\pi }{2}$,
the constant of integration $k\in \left[ 0,1\right] $. In contrast, for $L>%
\frac{\pi }{2}$, the allowed values for $k$ are $k=0$ and $k\in \left[
k_{m},1\right] $, where $k_{m}$ is determined from the condition
\begin{equation}
K\left( k_{m}\right) =L.  \label{3.7}
\end{equation}%
Condition (\ref{3.5}) yields an equation for $y_{0}$,%
\[
sn\,\left( L,k\right) dn\,\left( L,k\right) sn\,\left( y_{0},k\right)
dn\,\left( y_{0},k\right) =0
\]%
(where $dn\,u=\frac{d}{du}am\,u$), whose unique solution is $y_{0}=0$. Thus,
solutions of type I, compatible with boundary conditions (\ref{1.6}), have
the form
\begin{equation}
\phi \left( y\right) =\pi \left( 2n+1\right) +2\arcsin \left[ k\,sn\left(
y,k\right) \right] ,\quad n=0,\pm 1,\ldots ,  \label{3.8}
\end{equation}%
where $k\in \left[ 0,1\right] $ for $L\leq \frac{\pi }{2}$, and $k=0$, $k\in %
\left[ k_{m},1\right] $ for $L>\frac{\pi }{2}$.

An analysis of stability of solutions (\ref{3.8}) is straightforward. Thus,
for $k=0$ we have
\begin{equation}
\phi \left( y\right) =\pi \left( 2n+1\right) ,\quad n=0,\pm 1,\ldots .
\label{3.9}
\end{equation}%
The exact eigenfunction in (\ref{2.16}) is $\psi _{0}=const>0$, which
immediately yields $\mu _{0}=-1$. For $k\neq 0$, the denominator in (\ref%
{2.15}) is positive, whereas%
\[
\frac{d^{2}\phi }{dy^{2}}\left( L\right) \equiv -sn\,\left( L,k\right)
dn\,\left( L,k\right) <0,\quad \frac{d^{2}\phi }{dy^{2}}\left( -L\right)
\equiv sn\,\left( L,k\right) dn\,\left( L,k\right) >0.
\]%
Therefore, $\mu _{0}<0$. In other words, solutions (\ref{3.8}) correspond to
saddle points of (\ref{1.1}) and, hence, are unstable and physically
unobservable.

\subsection{Solutions of type II}

Consider solutions of type II [Eqs. (\ref{3.3})]. In contrast to the
solutions of type I [see (\ref{3.5.1})], now
\begin{equation}  \label{3.9.1}
\left| \phi \left( L\right) -\phi \left( -L\right) \right| <\infty
\end{equation}
for any $0<L<\infty $. Using the derivative
\begin{equation}  \label{3.10}
\frac{d\phi _{\pm }}{dy}=\pm \frac 2k\,dn\,\left( \frac{y-y_0}k,k\right) ,
\end{equation}
from $\,$(\ref{3.4}) we conclude that the appropriate solution is $\phi
=\phi _{+}$, with $k\in \left( 0,1\right] $. Condition (\ref{3.5}) yields an
equation for $y_0$,
\begin{equation}  \label{3.11}
sn\,\left( \frac Lk,k\right) cn\,\left( \frac Lk,k\right) sn\,\left( \frac{%
y_0}k,k\right) cn\,\left( \frac{y_0}k,k\right) =0.
\end{equation}
If $\frac Lk\neq pK\left( k\right) $ ($p=1,2,\ldots $) in (\ref{3.11}), this
equation has two different solutions:
\begin{equation}  \label{3.12}
y_0=-kK\left( k\right) ,\quad y_0=0.
\end{equation}
Correspondingly, solutions (\ref{3.3}), compatible with boundary conditions (%
\ref{1.6}), split into two distinct sets:
\begin{equation}  \label{3.13}
\phi _e\left( y\right) =\pi \left( 2n+1\right) +2\,am\left( \frac yk+K\left(
k\right) ,k\right) ,\quad n=0,\pm 1,\ldots ,
\end{equation}
and
\begin{equation}  \label{3.14}
\phi _o\left( y\right) =\pi \left( 2n+1\right) +2\,am\left( \frac
yk,k\right) ,\quad n=0,\pm 1,\ldots .
\end{equation}
The meaning of the subscripts $e$ (even) and $o$ (odd) should be clear from
the following: for solutions (\ref{3.13}), we have $\phi _e\left( 0\right)
=2\pi \left( n+1\right) $ ($n=0,\pm 1,\ldots $); in contrast, for solutions (%
\ref{3.14}), $\phi _o\left( 0\right) =\pi \left( 2n+1\right) $ ($n=0,\pm
1,\ldots $). Using Eq. (\ref{1.5}) and its derivative,
\begin{equation}  \label{3.15}
\frac{d^3\phi }{dy^3}=\cos \phi \frac{d\phi }{dy},
\end{equation}
we find that the ''local magnetic field'' (\ref{1.3}) at $y=0$ has a minimum
for $\phi _e$ and a maximum for $\phi _o$.

It is interesting to note that the two sets of solutions (\ref{3.13}), (\ref%
{3.14}) are related by the B\"acklund transformations

\[
\frac 12\frac d{dy}\left[ \phi _o\pm \phi _e\right] =\frac{1\pm \sqrt{1-k^2}}%
k\sin \frac{\phi _o\mp \phi _e}2,
\]
which, in the general case of a time-dependent sine-Gordon equation, is a
hallmark of complete integrability.\cite{L80,DEGM82,KK89} By virtue of the
symmetry relations%
\[
\phi _e\left( -y\right) =-\phi _e\left( y\right) +4\pi \left( n+1\right) ,
\]
\[
\phi _o\left( -y\right) =-\phi _o\left( y\right) +2\pi \left( 2n+1\right) ,
\]
the eigenfunction $\psi _0$ in (\ref{2.8}), (\ref{2.9}) is necessarily
symmetric with respect to reflection: $\psi _0\left( -y\right) =\psi
_0\left( y\right) $. As a result, for $\delta \phi \equiv const\,\psi _0$,
all the odd terms $\delta ^{2m+1}\Omega _G$ ($m=1,2,\ldots $) in expansion (%
\ref{2.2}) vanish by symmetry.

Stability regions for solutions (\ref{3.13}) and (\ref{3.14}) can be
established as follows. First, we note that the roots of the equations
\begin{equation}  \label{3.16}
pkK\left( k\right) =L,\quad p=1,2,\ldots ,
\end{equation}
$k=k_p\in \left( 0,1\right] $, form an infinite decreasing sequence of
bifurcation points:
\begin{equation}  \label{3.17}
1>k_1>k_2>\ldots
\end{equation}
Indeed, the second derivatives of $\phi _e$ and $\phi _o$, respectively,
read:
\begin{equation}  \label{3.18}
\frac{d^2\phi _e}{dy^2}\left( y\right) =2\sqrt{1-k^2}\frac{sn\,\left( \frac
yk,k\right) cn\,\left( \frac yk,k\right) }{dn^2\,\left( \frac yk,k\right) }%
,\quad \frac{d^2\phi _o}{dy^2}\left( y\right) =-2sn\,\left( \frac
yk,k\right) cn\,\left( \frac yk,k\right) .
\end{equation}
Using (\ref{3.18}), we find
\begin{equation}  \label{3.19}
\frac{d^2\phi _{e,o}}{dy^2}\left( \pm L\right) =0
\end{equation}
for $k=k_p$ ($p=1,2,\ldots $). Substituting $\phi =\phi _{e,o}$ with $k=k_p$
($p=1,2,\ldots $) into (\ref{2.15}), using (\ref{3.19}) and the fact that
the denominator in (\ref{2.15}) is strictly positive, we find $\mu _0=0$
both for $\phi =\phi _e$ and $\phi =\phi _o$. [As a matter of fact, the
derivatives $\frac{d\phi _{e,o}}{dy}$, for $k=k_p$ ($p=1,2,\ldots $),
coincide (up to a normalization factor) with the eigenfunctions $\psi
_{0\,e,o}$ corresponding to the eigenvalues $\mu _{0\,e,o}=0$.]

The bifurcation points $k=k_{p}$ ($p=1,2,\ldots $) subdivide the interval $%
I\equiv \left( 0,1\right] $ into an infinite set of semi-open intervals:
\begin{equation}
I=\cup _{p=0}^{\infty }I_{p},  \label{3.20}
\end{equation}%
where
\begin{equation}
I_{0}=\left( k_{1},1\right] ;\quad I_{p}=\left( k_{p+1},k_{p}\right] ,\quad
p=1,2,\ldots .  \label{3.21}
\end{equation}%
The index of these intervals $p=0,1,2,\ldots $ can be expressed as
\begin{equation}
p=\left[ \frac{\phi _{e,o}\left( L\right) -\phi _{e,o}\left( -L\right) }{%
2\pi }\right] ,  \label{3.22}
\end{equation}%
where $\left[ \ldots \right] $ stands for the integer part of the argument.
As can be easily verified using (\ref{2.15}) and (\ref{3.18}), $\mu _{0\,e}>0
$ and $\mu _{0\,o}<0$ for $k\in I_{0}$; $\mu _{0\,e}\geq 0$ and $\mu
_{0\,o}\leq 0$ for $k\in I_{2m}$ ($m=1,2,\ldots $); $\mu _{0\,e}\leq 0$ and $%
\mu _{0\,o}\geq 0$ for $k\in I_{2m+1}$ ($m=0,1,2,\ldots $). [We again
emphasize that the equalities $\mu _{0\,e,o}=0$ are realized only at the
bifurcation points $k=k_{p}$ ($p=1,2,\ldots $).] On these grounds, we
conclude that stability regions for the solutions $\phi =\phi _{e}$ are
given by the intervals $I_{2m}$ ($m=0,1,2,\ldots $), whereas stability
regions for the solutions $\phi =\phi _{o}$ are given by the intervals $%
I_{2m+1}$ ($m=0,1,2,\ldots $).

It is instructive to illustrate the above general results by explicit
evaluation of $\mu _0=\mu _0\left( k\right) $ for $k\ll 1$. In this limit,
the lowest eigenvalue $\mu _0$ can be expanded in a power series of $k$:%
\[
\mu _0=\sum_{n\geq 1}\mu _0^{\left( n\right) }\left( k\right) ,
\]
where $\mu _0^{\left( n\right) }\left( k\right) $ is of order $k^n$ ($%
n=1,2,\ldots $). Here, we restrict ourselves to the evaluation of $\mu
_0^{\left( 1\right) }\left( k\right) $: i.e., $\mu _0\approx \mu _0^{\left(
1\right) }\left( k\right) $. In view of the asymptotic expansions
\begin{equation}  \label{3.23}
\phi _e\left( y\right) \approx 2\pi \left( n+1\right) +\frac{2y}k-\frac{ky}2-%
\frac{k^2}4\sin \frac{2y}k,\quad n=0,\pm 1,\ldots ,
\end{equation}
\begin{equation}  \label{3.24}
\phi _o\left( y\right) \approx \pi \left( 2n+1\right) +\frac{2y}k-\frac{ky}2+%
\frac{k^2}4\sin \frac{2y}k,\quad n=0,\pm 1,\ldots .,
\end{equation}
it is sufficient to take a zeroth-order approximation to $\psi _0$: $\psi
_0\approx \psi _0^{\left( 0\right) }$. In zeroth order in $k$, equations (%
\ref{2.8}), (\ref{2.9}) for $\psi _0$ become%
\[
\frac{d^2\psi _0^{\left( 0\right) }}{dy^2}=0,\quad \frac{d\psi _0^{\left(
0\right) }}{dy}\left( -L\right) =\frac{d\psi _0^{\left( 0\right) }}{dy}%
\left( L\right) =0,
\]
whose solution is
\begin{equation}  \label{3.25}
\psi _{0\,e,o}^{\left( 0\right) }=const.
\end{equation}
Upon substitution of (\ref{3.23})-(\ref{3.25}) into (\ref{2.16}), we obtain
\begin{equation}  \label{3.26}
\mu _{0\,e}^{\left( 1\right) }=k\sin \frac Wk,\quad \mu _{0\,o}^{\left(
1\right) }=-k\sin \frac Wk.
\end{equation}
Expressions (\ref{3.26}) immediately yield bifurcation points for $k\ll 1$:
\begin{equation}  \label{3.27}
k_p=\frac W{\pi p},\quad p=1,2,\ldots ,
\end{equation}
in full agreement with (\ref{3.16}).

Summarizing, in this section we have derived a complete set of the solutions
to (\ref{1.5}) (both stable and unstable), compatible with boundary
conditions (\ref{1.6}). The only stable solutions [i.e., corresponding to
the minima of (\ref{1.1})] are those given by Eq. (\ref{3.13}) and Eq. (\ref%
{3.14}), where $k\in I_{2m}$ ($m=0,1,2,\ldots $) and $k\in I_{2m+1}$ ($%
m=0,1,2,\ldots $), respectively. In the next two sections both the stable
and unstable solutions will be analyzed in more detail.

\section{Stable Meissner and vortex (soliton) solutions}

It is convenient to introduce a unified classification of the stable
solutions, directly related to their physical interpretation. To this end,
we introduce a new integer, the vortex (or topological) number $%
N_{v}=0,1,\ldots $, by means of the definition
\begin{equation}
N_{v}\equiv p=\left[ \frac{\phi \left( L\right) -\phi \left( -L\right) }{%
2\pi }\right] =\left[ \frac{1}{2\pi }\int\limits_{-L}^{L}dy\frac{d\phi }{dy}%
\right]   \label{4.1}
\end{equation}%
[compare Eq. (\ref{3.22})], and fix the so far arbitrary integer $n=0,\pm
1,\pm 2,\ldots $ in Eqs. (\ref{3.13}) and (\ref{3.14}) by the condition
\begin{equation}
\phi \left( 0\right) =\pi N_{v}.  \label{4.2}
\end{equation}%
[Here, $\phi =\phi _{e}$ for $N_{v}=2m$ ($m=0,1,2,\ldots $), and $\phi =\phi
_{o}$ for $N_{v}=2m+1$ ($m=0,1,2,\ldots $).] Finally, using boundary
conditions (\ref{1.6}) for the determination of $k$, we arrive at the
desired form for the stable solutions:\cite{K04,K05}
\begin{equation}
\phi _{e}(y)=\pi \left( N_{v}-1\right) +2\,am\,\left( \frac{y}{k}+K\left(
k\right) ,k\right) ,\quad k=k\left( H\right) :  \label{4.3}
\end{equation}%
\begin{equation}
dn\,\left( \frac{L}{k},k\right) =\frac{\sqrt{1-k^{2}}}{kH},\qquad
N_{v}=2m\quad \left( m=0,1,\ldots \right) ;  \label{4.4}
\end{equation}%
\begin{equation}
\phi _{o}(y)=\pi N_{v}+2\,am\,\left( \frac{y}{k},k\right) ,\quad k=k\left(
H\right) :  \label{4.5}
\end{equation}%
\begin{equation}
dn\,\left( \frac{L}{k},k\right) =kH,\qquad N_{v}=2m+1\quad \left(
m=0,1,\ldots \right) ,  \label{4.6}
\end{equation}%
where $N_{v}=0$\ corresponds to the vortex-free Meissner (\textquotedblright
vacuum\textquotedblright ) solution, and $N_{v}=1,2\ldots $\ correspond to
vortex (soliton) solutions. The stability regions in terms of the field $H$
take the form
\begin{equation}
0\leq H<H_{0},\quad N_{v}=0,  \label{4.7}
\end{equation}%
\begin{equation}
\sqrt{H_{N_{v}-1}^{2}-1}\leq H<H_{N_{v}},\quad N_{v}=1,2,\ldots ,
\label{4.8}
\end{equation}%
with $H_{N_{v}}$ implicitly determined by
\begin{equation}
\left( N_{v}+1\right) K\left( \frac{1}{H_{N_{v}}}\right) =H_{N_{v}}L.
\label{4.9}
\end{equation}%
According to the results of section IV, we have $\delta ^{2}\Omega _{G}>0$
within the whole semi-open interval (\ref{4.7}). Analogously, $\delta
^{2}\Omega _{G}>0$ inside the semi-open intervals (\ref{4.8}), whereas $%
\delta ^{2}\Omega _{G}\geq 0$ at their boundaries. [Note that the upper
bounds of the stability regions (\ref{4.7}), (\ref{4.8}) are also determined
by the condition $\delta ^{2}\Omega _{G}\geq 0$.]

Solutions (\ref{4.3})-(\ref{4.9}) satisfy the symmetry relations
\begin{equation}  \label{4.9.1}
\phi \left( -y\right) =-\phi \left( y\right) +2\pi N_v
\end{equation}
and obey the boundary conditions
\begin{equation}  \label{4.9.2}
-\pi <\phi \left( -L\right) \leq 0,\quad 2\pi N_v\leq \phi \left( L\right)
<2\pi \left( N_v+\frac 12\right) ,
\end{equation}
which ensures the fulfillment of the stability conditions
\begin{equation}  \label{4.9.3}
\frac{d^2\phi }{dy^2}\left( -L\right) \leq 0,\quad \frac{d^2\phi }{dy^2}%
\left( L\right) \geq 0.
\end{equation}
As to major properties of the stable solutions, we want to emphasize the
following.

The stable solutions form an infinite set and their existence regions (\ref%
{4.7}), (\ref{4.8}) overlap (at least for two neighboring states). This fact
not only ensures that the stable solutions cover the whole field range $%
0\leq H<\infty $, but also proves that hysteresis is an intrinsic property
of any Josephson junction, irrespectively of the value of $W<\infty $.
However, overlapping is stronger for larger values of $W$ and at $1\ll
W<\infty $ may involve several neighboring states. On the other hand,
overlapping decreases with an increase of $H$.

Soliton (vortex) solutions with $N_v=1,2,\ldots $ exist for arbitrary small $%
W<\infty $, provided the field $H>0$ is sufficiently high. We also note that
the single-soliton ($N_v=1$) solution appears at a finite (for any $W<\infty
$) field $H=\sqrt{H_0^2-1}>0$ , which should be contrasted with the case of
the infinite interval $\left( -\infty ,+\infty \right) $: see Eq. (\ref{4.11}%
) below.

Solutions with $N_v=1,2,\ldots $are pure solitons only at $H=\sqrt{%
H_{N_v-1}^2-1}$, when $\phi \left( -L\right) =0$, $\phi \left( L\right)
=2\pi N_v$ and $\delta ^2\Omega _G\geq 0$. In the rest of the stability
regions (\ref{4.8}), when $\delta ^2\Omega _G>0$, we have solitons
''dressed'' by the Meissner field. As a matter of fact, solitons (vortices)
are confined to the spatial interval $\left[ -l,l\right] $, where $l$ is
determined from the conditions $\phi \left( -l\right) =0$, $\phi \left(
l\right) =2\pi N_v$. In contrast, the intervals $\left[ -L,-l\right) $ and $%
\left( l,L\right] $ are ''reserved'' for the Meissner field. Nevertheless,
although Josephson vortices always exist against a background of the
Meissner field, solutions (\ref{4.3})-(\ref{4.9}) with $N_v=1,2,\ldots $ by
no means can be thought of as a mere superposition of the Meissner and the
vortex fields, because the principle of superposition does not hold for the
non-linear equation (\ref{2.5}).

A solution with $N_v=0,1,2,\ldots $ solitons cannot be continuously
transformed into the solution with $N_v+1$ solitons and vice versa.
Therefore, any transitions between configurations with different vortex
numbers $N_v$ are necessarily thermodynamic first-order phase transitions.

Some examples of the stable solutions (for $N_v=0,1,2$), obtained by
numerical evaluation of (\ref{4.3})-(\ref{4.9}), are presented in Fig. 2. In
several particular cases, known from the previous literature, solutions (\ref%
{4.3})-(\ref{4.9}) can be expressed in terms of elementary functions. For
example, at $H=0$ and arbitrary $L<\infty $, there exists only the trivial
Meissner ($N_v=0$) solution $\phi _0\left( y\right) \equiv 0$, $y\in \left[
-L,L\right] $. In the low-field limit $0\leq H\ll 1$, the Meissner solution
[equation (\ref{4.3}) with $N_v=0$] reads:%
\[
\phi _0\left( y\right) \approx \frac{2H}{\cosh L}\sinh y,\quad h\left(
y\right) \approx \frac H{\cosh L}\cosh y.
\]
\begin{figure}[tbp]
\includegraphics{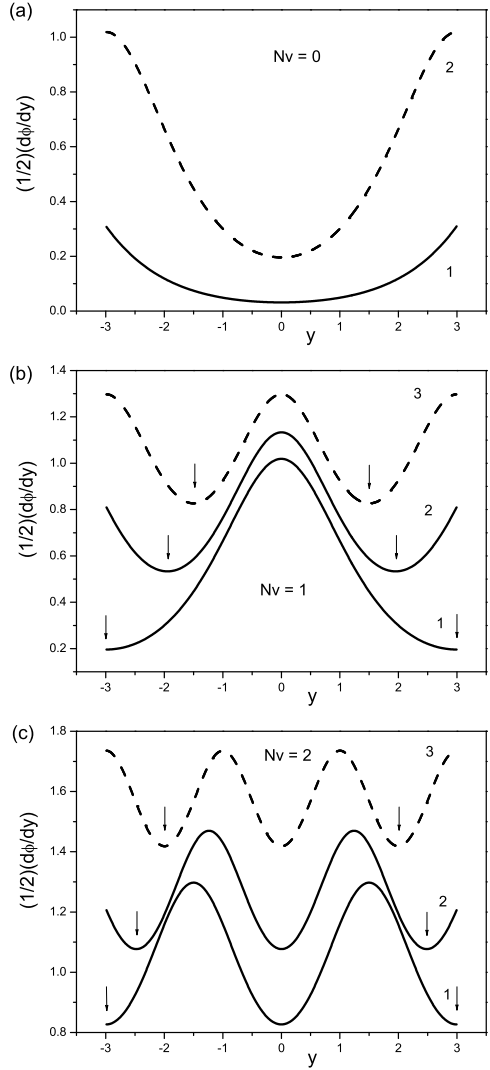}
\caption{Stable solutions for $L=3.0$. (a) The Meissner solution ($N_v=0$):
curves 1 and 2 correspond to the cases $0<H<H_0$ and $H=H_0$, respectively.
(b) The single-vortex solution ($N_v=1$): curves 1-3 correspond to the cases
$H=\protect\sqrt{H_0^2-1}$, $\protect\sqrt{H_0^2-1}<H<H_1$ and $H=H_1 $,
respectively; the vortex is confined to the spatial intervals denoted by
vertical arrows. (c) The two-vortex solution ($N_v=2$): curves 1-3
correspond to the cases $H=\protect\sqrt{H_1^2-1}$, $\protect\sqrt{H_1^2-1}%
<H<H_2$ and $H=H_2 $, respectively; the vortices are confined to the spatial
intervals denoted by vertical arrows.}
\label{fig:stable}
\end{figure}

By changing the variable $y\rightarrow y-L$ and proceeding to the limit $%
L\rightarrow \infty $ in equations (\ref{4.3}), (\ref{4.4}) with $N_v=0$, we
obtain the exact Meissner solution in a semiinfinite interval:\cite{KY72}
\begin{equation}  \label{4.10}
\phi _{0\,\infty }(y)=-4\arctan \frac{H\exp \left( -y\right) }{1+ \sqrt{1-H^2%
}},\quad y\in [0,+\infty ).
\end{equation}
Equation (\ref{4.10}) immediately yields the upper bound of the existence of
the Meissner state (or the superheating field of the Meissner state) in the
semiinfinite junction: $H_0\equiv H_s=1$. The local magnetic field induced
by (\ref{4.10}),
\begin{equation}  \label{4.10.1}
h\left( y\right) =\frac{H\left( 1+\sqrt{1-H^2}\right) }{\left( 1+\sqrt{1-H^2}%
\right) \exp y-H^2\sinh y},\quad y\in [0,+\infty ),
\end{equation}
vanishes at $y=+\infty $, as it should.

By proceeding to the limit $\pm $$L\rightarrow \pm \infty $, $H\rightarrow 0$%
, we find two different solutions: namely, the trivial Meissner ($N_v=0$)
solution $\phi _0\left( y\right) \equiv 0$, $y\in \left( -\infty ,+\infty
\right) $, and the well-known single-soliton ($N_v=1$) solution in an
infinite interval\cite{L80,DEGM82,KK89}
\begin{equation}  \label{4.11}
\phi _{1\,\infty }(y)=4\arctan \exp y,\quad y\in (-\infty ,+\infty ).
\end{equation}
The local magnetic field induced by (\ref{4.11}),
\begin{equation}  \label{4.11.1}
h\left( y\right) =\cosh {}^{-1}y,\quad y\in (-\infty ,+\infty ),
\end{equation}
vanishes at $y=\pm \infty $. Note that, from a point of view of the analysis
of stability, solutions (\ref{4.10}), (\ref{4.11}) constitute singular
cases: see appendix A.

Of particular interest is the limit $H\gg \max \left\{ 1,W\right\} $, which,
physically, corresponds to negligibly small screening by Josephson currents.
In this limit, the overlapping of states with different $N_v$ practically
vanishes, and equations (\ref{4.4})-(\ref{4.6}) become\cite{K99}%
\begin{equation}  \label{4.12}
\phi _{e,o}(y)\approx \pi N_v+2Hy -\frac{\left( -1\right) ^{N_v}}{4H^2}\left[
\sin \left( 2Hy\right) -2Hy\cos \left( HW\right) \right] ,
\end{equation}
where $N_v=\left[ \frac{HW}\pi \right] $, according to (\ref{4.1}). The
distribution of the local magnetic field induced by (\ref{4.12}) is
\begin{equation}  \label{4.12.1}
h\left( y\right) \approx H-\frac{\left( -1\right) ^{N_v}}{4H}\left[ \cos
\left( 2Hy\right) -\cos \left( HW\right) \right] .
\end{equation}
Equations (\ref{4.12}), (\ref{4.12.1}) explicitly demonstrate the existence
of solitons (or Josephson vortices) in the case $W\ll 1$.

Upon substitution of (\ref{4.3}), (\ref{4.4}) into (\ref{1.1}), we obtain
exact, closed-form analytical expressions for the Gibbs free energy. It is
convenient to write down these expressions in terms of the average energy
density $\omega \left( H,k\right) \equiv \frac{\Omega _{G}\left( H,k\right)
}{W}$:

\begin{equation}  \label{4.13}
\omega \left( H,k\right) =\omega _e\left( H,k\right) \delta _{N_v,2m}+\omega
_o\left( H,k\right) \delta _{N_v,2m+1}\quad \left( m=0,1,\ldots \right) ,
\end{equation}
\begin{eqnarray}  \label{4.14}
\omega _e\left( H,k\right) =2H^2+\frac 8W\left[ \frac 1kE\left( \frac
W{2k}+K\left( k\right) ,k\right) -\frac 1kE\left( k\right) \right. \left. -%
\frac{\left( 1-k^2\right) W}{4k^2}\right]  \nonumber \\
-\frac{8H}W\left[ am\,\left( \frac W{2k}+K\left( k\right) ,k\right) -\frac
\pi 2\right] ,\quad k=k\left( H\right);  \nonumber \\
\omega _o\left( H,k\right) =2H^2+\frac 8W\left[ \frac 1kE\left( \frac
W{2k},k\right) -\frac{\left( 1-k^2\right)W}{4k^2}\right]  \nonumber \\
-\frac{8H}W am\,\left( \frac W{2k},k\right) ,\quad k=k\left( H\right) ,
\end{eqnarray}
where $\delta _{N_v,2m}$ and $\delta _{N_v,2m+1}$ are the Kronecker indices;
$E\left( u,k\right) $ and $E\left( k\right) $\ are, respectively, the
incomplete and complete elliptic integrals of the second kind.\cite{AS65}
Note that in the limit $H\gg \max \left\{ 1,W\right\} $ [i.e., in the domain
of validity of (\ref{4.12})], equations (\ref{4.13}), (\ref{4.14}) reduce to
\begin{equation}  \label{4.15}
\omega \left( H\right) \approx 1-\frac{\left| \sin \left( HW\right) \right|
}{HW}+\frac 1{8H^2}\left[ \cos {}^2\left( HW\right) +\frac 12\right] .
\end{equation}

In Figs. 3(a) and 3(b), we present $\omega \left( H\right) $, obtained by
numerical evaluation of (\ref{4.13}), (\ref{4.14}), for the cases of a
''large'' ($L=3$) junction and a ''small'' ($L=0.3$) junction, respectively.
Figure 3(a) exhibits strong overlapping of neighboring stable
configurations, whereas in Fig. 3(b) overlapping is practically invisible.
The envelope of the energy curves for $N_v=0,1,\ldots ,6$ corresponds to the
absolute minimum of the Gibbs free energy at a given $H$ (a
thermodynamically stable configuration). Parts of the energy curves that lie
above the envelope in Fig. 3(a) correspond to local minima of the Gibbs free
energy (thermodynamically metastable configurations). For better orientation
in the physical situation, we have specified the upper bound of the
existence of the Meissner state ($H=H_0$) and the first thermodynamic
critical field\cite{J65,KY72,BP82} ($H=H_{c1}$). (The latter field is
determined by the requirement that the Gibbs free energies of the states $%
N_v=0$ and $N_v=1$ be equal to each other.) Both the fields, $H_0$ and $%
H_{c1}$, strongly depend on the length $W$: they increase (although at
different rates) with a decrease of $W$. For example, for $1\ll W<\infty $,
they are approximately given by $H_0\approx H_s=1$ and $H_{c1}\approx \frac
2\pi $, whereas for $W\ll 1$ they practically coincide: $H_{c1}\approx
H_0\approx \frac \pi W$ [see Fig. 3(b)]. Note that by decreasing the
external field below $H=H_{c1}$, one can still observe the single-vortex
state down to the field $H=\sqrt{H_0^2-1}<H_{c1}$ [the abscissa of the left
end of the energy curve for $N_v=1$ in Fig. 3(a)].
\begin{figure}[tbp]
\includegraphics{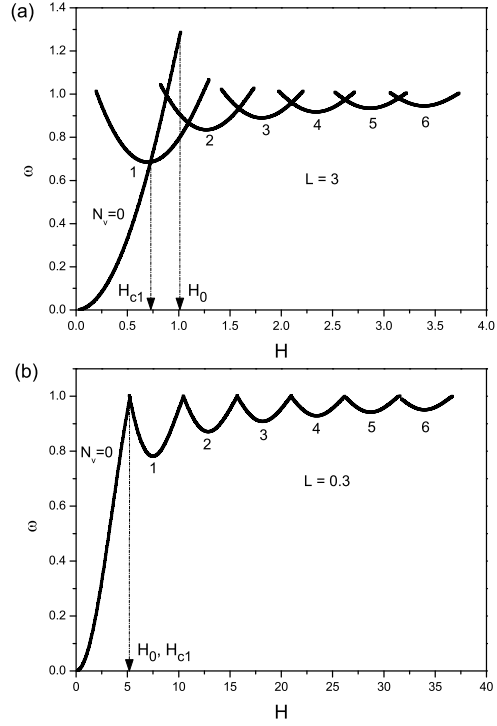}
\caption{The average Gibbs free-energy density $\protect\omega \left(
H\right) \equiv \frac{\Omega _G\left( H\right) }W$ for the cases of a
''large'' ($L=3.0$) junction (a) and a ''small'' ($L=0.3$) junction (b).
Energy curves are shown for $N_v=0,1,\ldots ,6$. The upper bound of the
existence of the Meissner state ($H_0$) and the first thermodynamic critical
field ($H_{c1}$) are also shown.}
\label{fig:energy}
\end{figure}

\section{Unstable (saddle-point) solutions}

There are three different types of unstable (saddle-point) solutions to (\ref%
{1.5}), (\ref{1.6}). Unstable solutions of the first and the second types
obey symmetry relations
\begin{equation}  \label{5.0}
\phi \left( -y\right) =-\phi \left( y\right) +2\pi Z,
\end{equation}
where $Z$ is an integer. In contrast to the stable solutions of section V,
this integer cannot be made to satisfy relation%
\[
Z=\left[ \frac{\phi \left( L\right) -\phi \left( -L\right) }{2\pi }\right]
\]
by any transformation $\phi \rightarrow \phi +2\pi n$ ($n=0,\pm 1,\ldots $)
and, thus, has no meaning of a topological number. In view of (\ref{5.0}),
the eigenfunction $\psi _0$ in (\ref{2.8}), (\ref{2.9}) is symmetric with
respect to reflection: $\psi _0\left( -y\right) =\psi _0\left( y\right) $.
However, in contrast to the stable solutions (\ref{4.3})-(\ref{4.6}), the
first two types of unstable solutions are characterized by the property
\begin{equation}  \label{5.3}
\frac{d^2\phi }{dy^2}\left( L\right) <0,\quad \frac{d^2\phi }{dy^2}\left(
-L\right) >0,
\end{equation}
which results in $\mu _0<0$ [see (\ref{2.15})].

The first type of unstable solutions is represented by the set (\ref{3.8}).
Without loss of generality, it is sufficient to consider the case of $n=0$:
\begin{equation}  \label{5.1}
\phi \left( y\right) =\pi +2\arcsin \left[ k\,sn\,\left( y,k\right) \right]
,\quad k=k\left( H\right) :\quad k\,cn\,\left( L,k\right) =H.
\end{equation}
Solution (\ref{5.1}) exists in the field range $0\leq H\leq \cosh {}^{-1}L$.
At $H=\cosh {}^{-1}L$, it becomes
\begin{equation}  \label{5.2}
\phi (y)=4\arctan \exp y,\quad y\in \left[ -L,L\right] .
\end{equation}
We emphasize that (\ref{5.2}) is unstable for any $L<\infty $, which should
be contrasted with the stable single-vortex solution in an infinite
interval, given by Eq. (\ref{4.11}). If $L\leq \frac \pi 2$, solution (\ref%
{5.1}) at $H=0$ degenerates into $\phi =\pi $. In contrast, if $L>\frac \pi
2 $, aside from the solution $\phi =\pi $, at $H=0$ there exists a
non-trivial solution
\begin{equation}  \label{5.4}
\phi \left( y\right) =\pi +2\arcsin \left[ k_m\,sn\,\left( y,k_m\right) %
\right] ,
\end{equation}
where $k_m$ is implicitly determined by (\ref{3.7}). Solution (\ref{5.1}) is
presented in Fig. 4(a).
\begin{figure}[tbp]
\includegraphics{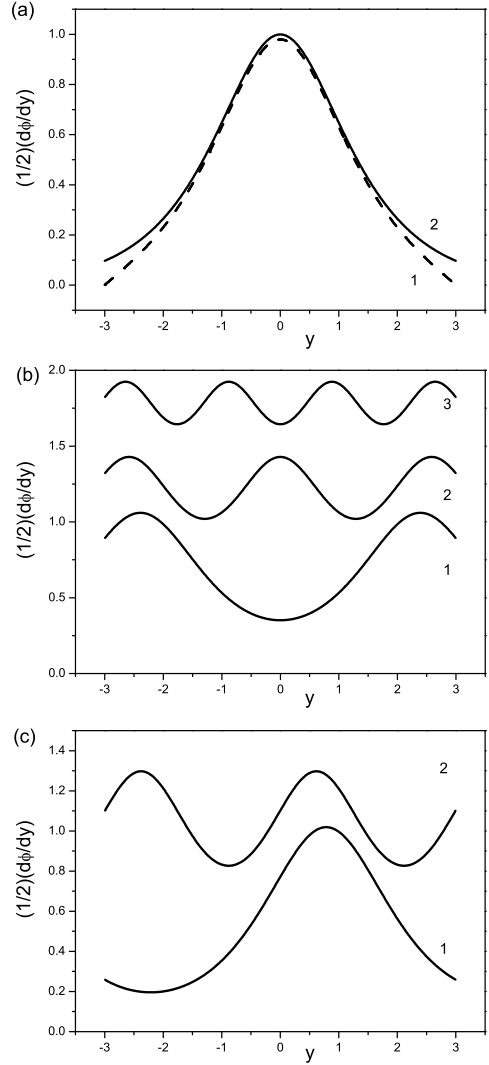}
\caption{Unstable (saddle-point) solutions for $L=3.0$. (a) The unstable
solution of the first type: curves 1 and 2 correspond to the cases $H=0$ and
$H=\cosh {}^{-1}L$, respectively. (b) Unstable solutions of the second type:
curves 1-3 correspond to different particular cases. (c) Unstable solutions
of the third type: curves 1 and 2 correspond to different particular cases.}
\label{fig:unstable}
\end{figure}

Unstable solutions of the second type can be obtained by prolonging
solutions (\ref{4.3})-(\ref{4.6}) beyond their respective stability regions (%
\ref{4.7}), (\ref{4.8}). Several examples of such solutions are given in
Fig. 4(b).

Finally, unstable solutions of the third type do not obey (\ref{5.0}).
Therefore, the eigenfunction $\psi _0$ in (\ref{2.8}), (\ref{2.9}) does not
possess reflection symmetry: $\psi _0\left( -y\right) \neq \psi _0\left(
y\right) $. These solutions can be obtained from solutions (\ref{3.13}), (%
\ref{3.14}), taken at the upper boundaries of their respective intervals of
stability, $I_{2m}$ with $m=1,2,\ldots $, and $I_{2m+1}$ with $%
m=0,1,2,\ldots $ (i.e., at the bifurcation points $k=k_p$, $p=1,2,\ldots $),
by a shift of the argument $y\rightarrow y-\alpha k$, where $\left| \alpha
\right| <K\left( k\right) $. This results in the properties
\begin{equation}  \label{5.5}
\frac{d\phi }{dy}\left( L\right) =\frac{d\phi }{dy}\left( -L\right) >0,\quad
\frac{d^2\phi }{dy^2}\left( L\right) =\frac{d^2\phi }{dy^2}\left( -L\right)
\neq 0.
\end{equation}
Relation (\ref{2.15}) yields $\mu _0<0$, because $\psi _0\left( L\right)
<\psi _0\left( -L\right) $ for $\frac{d^2\phi }{dy^2}\left( \pm L\right) >0$%
, and $\psi _0\left( L\right) >\psi _0\left( -L\right) $ for $\frac{d^2\phi
}{dy^2}\left( \pm L\right) <0$. Examples of unstable solutions of the third
type are presented in Fig. 4(c).

\section{Summary and conclusions}

The main results of this paper can be summarized as follows. Mathematically,
we have completely solved the ill-posed boundary value problem (\ref{1.5}), (%
\ref{1.6}) and presented an exhaustive classification of the obtained
solutions with respect to their stability. Our exact analytical treatment of
the issue of stability can be easily extended to include more difficult
cases, e.g., such as Eq. (\ref{1.5}) under more general boundary conditions%
\cite{OS67} or a coupled system of static sine-Gordon equations.\cite%
{K04,K05}

Physically, the complete set of stable solutions (\ref{4.3})-(\ref{4.9})
gives a clear final answer to the question what should be called the
Meissner state and the vortex structure in Josephson junctions for arbitrary
$W<\infty $ and $0\leq H<\infty $. We have shown that the stability of
physical solutions to (\ref{1.5}), (\ref{1.6}) is nothing but a consequence
of soliton boundary conditions (\ref{4.9.2}) [or, equivalently, (\ref{4.9.3}%
)]. In particular, the stability conditions (\ref{4.9.3}) provide a simple,
rigorous criterion to distinguish between physical (stable) and unphysical
(unstable) solutions, which can be readily verified by comparing Fig. 2 with
Fig. 4.

To illustrate the difference between the results of our paper and those of
previous publications, we turn to a two-parameter expression
\begin{equation}  \label{6.1}
\phi \left( y\right) =\pi +2\,am\,\left( \frac{y-y_0}k,k\right) .
\end{equation}
Proposed in the previous literature\cite{KY72,BP82,K66,OS67,So96} as an
''exact vortex solution'', expressions of the type (\ref{6.1}) by no means
can be regarded as such in the absence of any explicit definition of the
constants of integration $k$, $y_0$ and of stability regions. As we have
shown, aside from the Meissner solution [equations (\ref{4.3}), (\ref{4.4}),
(\ref{4.7}), (\ref{4.9}) with $N_v=0$] and the actual vortex solutions
[equations (\ref{4.3})-(\ref{4.9}) with $N_v=1,2,\ldots $], the
two-parameter family (\ref{6.1}) contains absolutely unstable, unobservable
solutions of the second and the third types [see section VI and Figs. 4(b),
4(c)].

The reader should be warned against confusion between the single-soliton
solution in the infinite interval $\phi _{1\,\infty }$ [equation (\ref{4.11}%
)] and its restriction onto a finite interval, the saddle-point solution (%
\ref{5.2}) [curve 2 in Fig. 4(a)]. Unfortunately, the unstable solution (\ref%
{5.2}) is often erroneously interpreted in literature\cite{Kr01} as a
''Josephson vortex''.

Historically, the single-soliton solution $\phi _{1\,\infty }$ served as the
first example of vortex solutions in Josephson structures.\cite{J65}
However, this solution cannot be realized experimentally in any realistic
(with $W<\infty $) physical system. In this regard, we want to emphasize
once again that the actual physical single-vortex solution is given by Eqs. (%
\ref{4.5}), (\ref{4.6}), (\ref{4.8}) and (\ref{4.9}) with $N_v=1$ [see also
Fig. 2(b)].

Concerning saddle-point solutions, classified and discussed in section VI,
they may play a certain role as channels of fluctuation-induced transitions
from thermodynamically metastable states [local minima of (\ref{1.1})] to a
thermodynamically stable state [an absolute minimum of (\ref{1.1})].
(Compare the decay of current-carrying states in narrow superconducting
channels.\cite{LA67}) However, this issue asks for further investigation
within the framework of a more general non-equilibrium approach.

Finally, as is pointed out in Refs. \cite{K04,K05} and in the Introduction,
the exact analytical solutions (\ref{4.3})-(\ref{4.9}) constitute a new
class of static topological solutions to the sine-Gordon equations. We
believe that they may find application not only in superconductivity, but
also in other fields of modern nonlinear physics that deal with sine-Gordon
equations.

\begin{center}
\bigskip\ \textbf{Acknowledgements}
\end{center}

We thank A. S. Kovalev and M. M. Bogdan for a discussion of solutions (\ref%
{4.3})-(\ref{4.9}) and of their possible application to some nonlinear
problems of mechanics and condensed-matter physics. We also thank V. A.
Marchenko, E. Ya. Khruslov, and V. P. Kotlyarov for a discussion of some
mathematical issues related to the new solutions.

\appendix

\section{{}{Analysis of stability in two singular cases}}

In this appendix, we present an analysis of stability of the Meissner
solution (\ref{4.10}) and the single-soliton solution (\ref{4.11}). This
analysis leads to singular Sturm-Liouville eigenvalue problems,\cite{LS70}
formulated on the intervals $\left[ 0,+\infty \right) $ and $\left( -\infty
,+\infty \right) $, respectively.

\subsection{Meissner solution in a semiinfinite interval}

The Meissner solution $\phi _{0\,\infty }$ [equation (\ref{4.10})] that
exists in the field range $0\leq H\leq 1$ is a stationary point of the
free-energy functional
\begin{equation}  \label{a1}
\Omega _{G\,\infty }\left[ \phi ,\frac{d\phi }{dy};H\right]
=\int\limits_0^\infty dy\left[ 1-\cos \phi \left( y\right) +\frac 12\left[
\frac{d\phi \left( y\right) }{dy}\right] ^2\right] -2H\phi \left( 0\right) .
\end{equation}
The second variation of (\ref{a1}) at $\phi =\phi _{0\,\infty }$ has the form%
\begin{equation}  \label{a2}
\delta ^2\Omega _{G\,\infty }\left[ \delta \phi ,\frac{d\delta \phi }{dy}%
\right] _{\phi =\phi _{0\,\infty }} =\int\limits_0^\infty dy\left[ \cos \phi
_{0\,\infty }\left( y\right) \left[ \delta \phi \left( y\right) \right] ^2+%
\left[ \frac{d\delta \phi }{dy}\left( y\right) \right] ^2\right] ,
\end{equation}
where the variation $\delta \phi $ has continuous first derivatives and
obeys boundary conditions
\begin{equation}  \label{a3}
\frac{d\delta \phi }{dy}\left( 0\right) =0,\quad \delta \phi \left( +\infty
\right) =0,\quad \frac{d\delta \phi }{dy}\left( +\infty \right) =0.
\end{equation}

At $H=0$, we have $\phi _{0\,\infty }\equiv 0$, and Eq. (\ref{a2})
immediately yields $\delta ^2\Omega _{G\,\infty }>0$ for all allowed
variations $\delta \phi $. To evaluate the sign of $\delta ^2\Omega
_{G\,\infty }$ in the interval $0<H\leq 1$, we have to evaluate the sign of
the lowest eigenvalue $\mu _0$ of the problem
\begin{equation}  \label{a4}
-\frac{d^2\psi }{dy^2}+\cos \phi _{0\,\infty }\left( y\right) \psi =\mu \psi
,\quad y\in \left[ 0,+\infty \right) ,
\end{equation}
\begin{equation}  \label{a5}
\frac{d\psi }{dy}\left( 0\right) =0,\quad \psi \left( +\infty \right) =0,
\end{equation}
where the normalizable eigenfunction $\psi _0$ has no nodes in the interval $%
\left[ 0,+\infty \right) $ and can be considered positive.

For $\mu _0$, we have the following general relation [compare with (\ref%
{2.15})]
\begin{equation}  \label{a6}
\mu _0=-\frac{\psi _0\left( 0\right) \frac{d^2\phi _{0\,\infty }}{dy^2}%
\left( 0\right) }{\int\limits_0^\infty dy\psi _0\left( y\right) \frac{d\phi
_{0\,\infty }}{dy}\left( y\right) }
\end{equation}
that holds in the whole interval $0<H\leq 1$. The evaluation of $\frac{%
d^2\phi _{0\,\infty }}{dy^2}\left( 0\right) $ yields:
\begin{equation}  \label{a7}
\frac{d^2\phi _{0\,\infty }}{dy^2}\left( 0\right) =-\frac{2H}{1+ \sqrt{1-H^2}%
}\left( 1+\sqrt{1-H^2}-H^2\right) .
\end{equation}
According to (\ref{a7}), $\frac{d^2\phi _{0\,\infty }}{dy^2}\left( 0\right)
<0$ for $0<H<1$, and $\frac{d^2\phi _{0\,\infty }}{dy^2}\left( 0\right) =0$
for $H=1$. Taking into account that the denominator in (\ref{a6}) is
positive [we remind that $\frac{d\phi _{0\,\infty }}{dy}\equiv 2h$], we
conclude that $\mu _0>0$ for $0<H<1$, and $\mu _0=0$ for $H=1$. Summarizing
the results for $H=0$ and $0<H\leq 1$, we state: $\delta ^2\Omega
_{G\,\infty }>0$ for $0\leq H<1$, and $\delta ^2\Omega _{G\,\infty }\geq 0$
for $H=1.$

\subsection{Single-soliton solution in the infinite interval}

The single-soliton solution $\phi _{1\,\infty }$ [equation (\ref{4.11})]
that exists at $H=0$ is a stationary point of the free-energy functional
\begin{equation}  \label{a8}
\Omega \left[ \phi ,\frac{d\phi }{dy}\right] =\int\limits_{-\infty }^\infty
dy\left[ 1-\cos \phi \left( y\right) +\frac 12\left[ \frac{d\phi \left(
y\right) }{dy}\right] ^2\right] .
\end{equation}
The standard analysis of $\delta ^2\Omega $ at $\phi =\phi _{1\,\infty }$
requires evaluation of the lowest eigenvalue $\mu _0$ of the problem
\begin{equation}  \label{a9}
-\frac{d^2\psi }{dy^2}+\cos \phi _{1\,\infty }\left( y\right) \psi =\mu \psi
,\quad y\in \left( -\infty ,+\infty \right) ,
\end{equation}
\begin{equation}  \label{a10}
\psi \left( \pm \infty \right) =0.
\end{equation}
As usual, the normalizable eigenfunction $\psi _0$ has no nodes in the
interval $\left( -\infty ,+\infty \right) $.

In this particular case, both $\mu _0$ and $\psi _0$ can be determined
explicitly. Indeed, the first derivative
\begin{equation}  \label{a11}
\chi \left( y\right) \equiv \frac{d\phi _{1\,\infty }}{dy}\left( y\right)
=\frac 2{\cosh y},\quad y\in \left( -\infty ,+\infty \right)
\end{equation}
has no nodes and satisfies boundary conditions $\chi \left( \pm \infty
\right) =0$. Moreover, it obeys the equation
\begin{equation}  \label{a12}
-\frac{d^2\chi }{dy^2}+\cos \phi _{1\,\infty }\left( y\right) \chi =0,\quad
y\in \left( -\infty ,+\infty \right) .
\end{equation}
Upon a comparison with (\ref{a9}), (\ref{a10}), we conclude that
\begin{equation}  \label{a13}
\psi _0\left( y\right) =const\,\cosh {}^{-1}y,\quad \mu _0=0.
\end{equation}

Thus, $\delta ^2\Omega \geq 0$ for $\phi =\phi _{1\,\infty }$, and the
single-soliton solution (\ref{4.11}) corresponds to a bifurcation state. In
view of translation symmetry of the problem, a shift of the argument $%
y\rightarrow y-y_0$, $\left| y_0\right| <\infty $ does not affect the
stability of the single-soliton solution $\phi _{1\,\infty }$, which should
be contrasted with the situation for soliton solutions in the bifurcation
state in the case of a finite interval $\left[ -L,L\right] $: see the last
paragraph of section VI and Fig. 4(c).


\begin{thebibliography}{99}
\bibitem{J65} B. D. Josephson, Advan. Phys. \textbf{14}, 419 (1965).

\bibitem{KY72} I. O. Kulik and I. K. Yanson, \textit{The Josephson Effect in
Superconductive Tunneling Structures} (Israel Program for Scientific
Translations, Jerusalem, 1972).

\bibitem{BP82} A. Barone and G. Paterno, \textit{Physics and Applications of
the Josephson Effect} (Wiley, New York, 1982).

\bibitem{K66} I. O. Kulik, Zh. Eksp. Teor. Fiz. \textbf{51}, 1952 (1966)
[Sov. Phys. JETP \textbf{24}, 1307 (1967)].

\bibitem{OS67} C. S. Owen and D. J. Scalapino, Phys. Rev. \textbf{164}, 538
(1967).

\bibitem{Ga84} Yu. S. Galperin and A. T. Filippov, Zh. Eksp. Teor. Fiz.
\textbf{86}, 1527 (1984) [Sov. Phys. JETP \textbf{59}, 894 (1984)].

\bibitem{Yu94} K. N. Yugay, N. V. Blinov, and I. V. Shirokov, Phys. Rev. B
\textbf{49}, 12036 (1994); Fiz. Nizk. Temp. \textbf{25}, 712 (1999).

\bibitem{Se04} E. G. Semerdjieva, T. L. Boyadjiev, and Yu. M. Shukrinov,
Fiz. Nizk. Temp. \textbf{30}, 610 (2004); Yu. M. Shukrinov, E. G.
Semerdjieva, and T. L. Boyadjiev, J. Low Temp. Phys. \textbf{139}, No.1
(2005).

\bibitem{BCG92} L. N. Bulaevskii, J. R. Clem, and L. I. Glazman, Phys. Rev.
B \textbf{46}, 350 (1992).

\bibitem{Kr01} V. M. Krasnov, Phys. Rev. B \textbf{63}, 064519 (2001).

\bibitem{CH} R. Curant and D. Hilbert, \textit{Methods of Mathematical
Physics} (Interscience, New York, 1962), Vol. II..

\bibitem{A70} N. I. Akhiezer, \textit{Elements of the Theory of Elliptic
Functions} (Nauka, Moscow, 1970) (in Russian)..

\bibitem{K04} S. V. Kuplevakhsky, Fiz. Nizk. Temp. \textbf{30}, 856 (2004)
[Low Temp. Phys. \textbf{30}, 646 (2004)].

\bibitem{K05} S. V. Kuplevakhsky, J. Low Temp. Phys. \textbf{139}, 141
(2005).

\bibitem{L80} G. R. Lamb Jr., \textit{Elements of Soliton Theory} (Wiley,
New York, 1980).

\bibitem{DEGM82} R. K. Dodd, J. C. Eilbeck, J. D. Gibbon, and H. C. Morris,
\textit{Solitons and Nonlinear Wave Equations} (Academic Press, London,
1982).

\bibitem{KK89} A. M. Kosevich and A. S. Kovalev, \textit{An Introduction to
Nonlinear Physical Mechanics} (Naukova Dumka, Kiev, 1989) (in Russian).

\bibitem{A62} N. I. Akhiezer, \textit{The Calculus of Variations} (Blaisdell
Publishing, New York, 1962).

\bibitem{Th82} J. M. T. Thompson, \textit{Instabilities and Catastrophes in
Science and Engineering} (Wiley, New York, 1982).

\bibitem{LL50} M. A. Lavrentiev and L. A. Liusternik, \textit{A Course of
the Calculus of Variations} (GITTL, Moscow, 1950) (in Russian).

\bibitem{r1} Relation (\ref{2.7}) can be readily verified, if one assumes
that $\delta $$\phi $ has continuous derivatives up to the second order and
uses the completeness of the set $\left\{ \psi _n\right\} _{n=0}^\infty $.
However, this relation holds even without the continuity of $\frac{d^2\delta
\phi }{dy^2}$.

\bibitem{LS70} B. M. Levitan and I. S. Sargsyan, \textit{An Introduction to
Spectral Theory} (Nauka, Moscow, 1970) (in Russian).

\bibitem{WW27} E. T. Whittaker and G. N. Watson, \textit{A Course of Modern
Analysis} (University Press, Cambridge, 1927).

\bibitem{AS65} M. Abramowitz and I. A. Stegun, \textit{Handbook of
Mathematical Functions} (Dover, New York, 1965).

\bibitem{K99} S. V. Kuplevakhsky, Phys. Rev. B \textbf{60}, 7496 (1999);
\textit{ibid.} \textbf{63}, 054508 (2001).

\bibitem{So96} S. N. Song, P.R. Auvil, M. Ulmer, and J. B. Ketterson, Phys.
Rev. B \textbf{53}, R6018 (1996).

\bibitem{LA67} J. S. Langer and V. Ambegaokar, Phys. Rev. \textbf{164}, 498
(1967).
\end{thebibliography}
\end{document}